\documentclass[aps,prd,twocolumn,showpacs]{revtex4}

\def\b{\bar}
\def\d{\partial}

\def\cD{{\cal D}}

\def\l{\lambda}

\def\m{\mu}
\def\n{\nu}

\def\q{\b q}

\def\~{\widetilde}

\def\bY3{\bar Y_{,3}}
\def\Y3{Y_{,3}}
\def\z{\zeta}
\def\Z{{\b\zeta}}
\def\Y{{\bar Y}}
\def\cZ{{\bar Z}}
\def\`{\dot}
\def\be{\begin{equation}}
\def\ee{\end{equation}}
\def\bea{\begin{eqnarray}}
\def\eea{\end{eqnarray}}

\def\cF{{\cal F}}

\def\mn{{\mu\nu}}

\begin{document}

\title{Kerr-Schild photonlike metric solutions}

\author{Alexander Burinskii}

\affiliation{Gravity Research Group, NSI Russian Academy of
Sciences, \\
B. Tulskaya 52  Moscow 115191 Russia;}

\begin{abstract}
The charged and spinning lightlike solutions are obtained and
analyzed in the Kerr-Schild formalism. One of them may be
considered as  ultrarelativistic boost of the Kerr-Newman solution
along the direction of angular momentum. The Kerr singular ring
disappears, going to infinity. However, there remains a finite
relativistic parameter $a$ which determines the twist of Kerr
congruence and total spin of the solutions by the Kerr relation
$J=ma$. In particular, assuming that this solution describes
gravitational field of a photon and setting $J=\hbar$ and $E=m$,
one obtains that $a$ is de Broglie wavelength. Electromagnetic
field of the solutions is aligned with the Kerr null congruence.
Some of the presented solutions contain singular lightlike beams.
\end{abstract}

\maketitle

\section{Introduction} The problem of finding the ultrarelativistic
limits of exact particle-like or BH- solutions of the Einstein
field equations have paid attention in the connection with some
non-trivial gravitational effects which are expected to occur in
the interparticle interactions at extreme energies due to the
presence of gravitational shock waves \cite{DratHo,FPVV} and
singular strings. Another field of application is  the study of
possible influence of gravitational field of photon.

First results in this directions were obtained by Aichelburg and
Sexl \cite{AicSex}, who considered the behavior of the
Schwarzschild metric under ultrarelativistic boost. The resulting
metric is a pp-wave, \cite{DratHo,FPV},

\be ds^2 = dudv + f(x,y)\delta(v) dv^2 +dx^2 + dy^2 \ee with a
front surface $v=0$ moving along the z axis. Here $v = (z -
t)2^{-1/2} $ and $u = (z + t)2^{-1/2}$. In particular, this
solution may also be considered as a metric of the Kerr-Schild
class \be g_\mn = \eta_\mn + 2H(v,x,y) k_\m k_\n ,\ee where $ k^\m
= dv = (dz - dt)2^{-1/2}$ is the expansion-free and twist-free
principal null congruence, and
 $\eta_\mn$ is metric of an auxiliary Minkowski space-time.
Similar treatments with ultrarelativistic boost of the Kerr
geometry have also been performed in many other works
\cite{FerPen,VegSan,LuoSan}, leading to the similar conclusion
that limiting metrics have the expansion-free and twist-free
pp-waves with an energy density $T_{vv}=\rho(x,y)\delta(v)$
distributed on singular  front. If the total mass-energy $m $
distributed on the
 front is finite, then the rest-mass $m_0$ of the corresponding
 stationary solution  has to tend to zero to provide the finite
 value for the relativistic mass-energy $E = m= m_0/\sqrt{1-(v/c)^2 }$
 in the limit $v\to c .$ Therefore, the limit $v\to c$ has to be
 related with the simultaneous limit $m_0 \to 0.$
\footnote{One can naively suppose that the limit $m_0\to 0$
becomes flat. Indeed, it differs from flat topologically. In
particular, in the Kerr case this limit may be performed with a
constant parameter
  $a ,$ leading to a space-time with twofold topology having the branch
 line along the naked Kerr singular ring. The corresponding
 Schwarzschild solution may be obtained by the subsequent limit
 $a\to 0,$ leading to two copies of Minkowski space glued at the
 singular point (line) $x=y=z=0 .$ Alternatively, changing the order
 of limits (doing $a=0$ first), one obtains one exemplar of the space
 with a punctured point.}

 If angular momentum of the Kerr solution is oriented along the boost
 direction, the Kerr singular ring will be orthogonal to the boost and
 will not be subjected to Lorentz contraction. During the limiting
 procedure its radius will be $a=a_0.$ Assuming that the rest mass of the
 prototype Kerr solution is infinitesimally small, i.e. $m_0\to 0$ by
 $v\to c ,$ and following the usual Kerr metric relation $ J=m_0 a_0 ,$
 we obtain that the limiting ultrarelativistic metric will have zero
 angular momentum, $\lim J |_{m_0\to 0} =0.$ We arrive at the conclusion
 that the most of the obtained ultrarelativistic analogs of the
 Kerr solution have the zero angular momentum.

However, one can also perform an alternative ultrarelativistic
limit of the Kerr solution, setting  $m_0\to 0,$ by the constant
value of the total angular momentum $J = m_0 a_0.$ Obviously, such
a limit has to be accompanied by simultaneous limit $a_0 \to
\infty , $ where $a_0$ is the radius of singular ring of the
corresponding Kerr solution at rest. In this case, one defines a
`relativistic' parameter $a=J/m$ and retains during the limit the
Kerr relation \be J=m_0 a_0 = m a \label{J} .\ee One sees that it
may be achieved setting $a= a_0 \sqrt{1-(v/c)^2}.$ Such a limit
for the uncharged Kerr solution was performed in
\cite{BurMag,BurMag1} by using the complex Kerr formalism, and
corresponding exact solution may also be obtained in the standard
Kerr-Schild formalism. This solution has a` finite length
parameter $a,$ however,  the Kerr singular ring disappears, going
to infinity in accordance with

\be a_0 = a/ \sqrt{1-(v/c)^2}. \label{a0}\ee

Recently, interest to the light-like solutions with angular
momentum was renewed, and there appears  series of the works on
the `gyraton' solutions, based on the Brinkman class of metrics
\cite{Fro} $ g_\mn = \eta_\mn + h_\mn ,$ which is relative to the
Kerr-Schild one, but has another type of the quasi-linear
contribution $h_\mn .$

In this paper we generalize the obtained earlier in
\cite{BurMag,BurMag1} solutions, incorporating electromagnetic
field which is aligned with respect to Kerr congruence. Note, that
in the gyraton solutions electromagnetic field is aligned with
respect to the Killing vector. In conclusion we discuss some new
problems which may be related to application of these solutions.
In particular, the problem of the obtaining more general solutions
with wave electromagnetic field which are related to case
$\gamma\ne 0$ of the Kerr-Schild formalism, and also based on the
Kerr theorem twistorial approach to interaction between the
lightlike and massive particles \cite{Multi}.

\section{Kerr-Schild formalism.} We follow to formalism and
notations of seminal work \cite{DKS}. Metric has the Kerr-Schild
form

\be g_\mn=\eta _\mn + 2h e^3_\m e^3_\n \label{gks} , \ee where
$\eta_\mn = diag \{-1,1,1,1 \}$ is the metric of auxiliary
Minkowski background $x^\m=(t,x,y,z)\in M^4 .$

The vector field $e^3_\m$ is tangent  to principal null congruence
which plays central role in the Kerr-Schild formalism, defining
the structure of all tensor quantities. It is described in the
 null Cartesian coordinates \bea
2^{1\over2}\z &=& x+iy ,\qquad 2^{1\over2} \Z = x-iy , \nonumber\\
2^{1\over2}u &=& z + t ,\qquad 2^{1\over2}v = z - t . \label{ncc}
\eea and has the form \be e^3= du + \bar Y d \zeta + Y d \bar\zeta
- Y \bar Y dv \label{e3} . \ee  The principal null congruence is
determined by the {\it Kerr theorem} in terms of the complex
function $Y\equiv Y(x^\m), \ x^\m\in M^4 .$ Besides the
congruence, function $Y(x^\m)$ determines a null tetrad. Direction
$e^3$ is completed to
 null tetrad $e^a, \ a=1,2,3,4 \ ,$
\be e^1=d \zeta - Y dv, \  e^2 = d \bar\zeta - \bar Y dv, \ e^4
=dv + h e^3. \label{ea} \ee

The null congruences (\ref{e3})  satisfying  the conditions \be
Y,_2=Y,_4 \label{Y24} ,\ee  are geodesic and shear-free. (Here
$,_a=e_a^\m\d_\m $ are the directional derivatives \footnote{In
explicit form they are
  $ \d_1 = \d_\z  - \Y \d_u ; \ \d_2 =  \d_\Z - Y \d_u ; \
 \d_3 =  \d_u - h \d_4 , \ $ and
 $ \d_4 =  \d_v + Y \d_\z + \Y \d_\Z - Y  \Y \d_u . $}).

 {\bf The Kerr theorem} claims that such congruences are determined by
 function $Y=Y(x^\m)$ which is a solution of the equation

\be F (Y,\l_1,\l_2) = 0 \ , \label{Fgen} \ee where  $F$ is an
arbitrary analytic function of the  projective twistor
coordinates

\be Y,\quad \l_1 = \z - Y v, \quad \l_2 =u + Y \Z \ . \label{Tw}
\ee

Integration of the Einstein-Maxwell field equations for the ansatz
(\ref{gks}) with a general geodesic and shear-free null field
$e^3_\m (x), \ x\in M^4$  was fulfilled in \cite{DKS}, leading to
the following form of the function $h :$

\be h = \frac 12 M(Z+\bar Z) - \frac 12 A \bar A Z\bar Z \ .
\label{h}\ee Therefore, the function $M,A, \bar A, Z$ and $\bar Z$
determine fully the metric, while the  strength tensor of
self-dual electromagnetic field $\cF _\mn =\cF _{ab}e^a e^b$ is
determined by the functions $A$ and $\gamma $ and has the
following nonzero tetrad components

\be \cF _{12} = \cF _{34}= AZ^2, \quad  \cF _{31} =\gamma Z -
(AZ),_1 \ . \label{Fab}\ee

The function $Z^{-1}$  characterizes a  radial distance which is
determined by the Kerr theorem and will be discussed later.
Differential equations for the unknown so far functions contain
{\bf electromagnetic sector:}

\be A,_2 - 2 Z^{-1} \cZ Y,_3 A = 0 , \quad A,_4=0 ,\label{E1}\ee

\be \cD A+ \cZ ^{-1} \gamma ,_2 - Z^{-1} Y,_3 \gamma =0, \quad
\gamma,_4=0 .\label{E2}\ee

Here \be \cD=\d _3 - Z^{-1} Y,_3 \d_1 - \cZ ^{-1} \Y ,_3 \d_2 .
\label{cD}\ee

{\bf Gravitational sector} contains two equations:

\bea M,_2 - 3 Z^{-1} \cZ Y,_3M = A\bar\gamma \cZ , \label{G5}  \\
\cD M = \frac 12 \gamma\bar\gamma  , \quad M,_4=0 . \label{G6}\eea

The necessary functions $Z,\ Y ,$ their directional derivatives
$,_a$ and parameters are determined by the generating function $F$
of the  Kerr theorem. In particular, the complex dilatation

\be Z= P/\tilde r \label{Z} \ee is related to `complex radial
distance'

\be\tilde r = - d F / d Y \ . \label{tr} \ee

Function $F(Y)$ used in \cite{DKS} has the general form

\be F \equiv \phi (Y)+ (q Y + c) \l_1 - (pY + \q) \l_2 \,
\label{F2} \ee where $\phi= a_0 +a_1 Y + a_2 Y^2 .$

The method developed in the papers \cite{BurMag,BurMag1} allows
one  to fix explicit values of these parameters corresponding to
concrete values of the boost and orientation of angular momentum.
The solutions of the main equation (\ref{Fgen}) can be found in
explicit form and correspond to the Kerr solution up to the
Lorentz boost, rotation, and shift of the origin.

Coefficients $ a_0,\quad a_1, \quad a_2 $ define orientation of
angular momentum and the constants $p, q, \q, c $ are related to
 Killing vector of the solution. They determine  function $ P = p
Y\bar Y +qY + \bar q \bar Y + c .$

In particular, for the Kerr-Newman solution at rest $p=c=2^{-1/2},
\quad q=\bar q=0$ and $P =2^{-1/2}(1+Y\Y).$ Spin $J$ is oriented
along z axis for $a_0=a_2=0 ,$ and $a_1=-ia .$

\bigskip

\section{Class of the light-like solutions.}
We consider the simplest case, taking the above parameters of the
Kerr solution at rest as a prototype of the boosted solution. We
have $\phi = -ia Y $ which corresponds to orientation of the
angular momentum in the z-direction, and we direct collinearly the
light-like boost which will be described by the parameters

\be p=q=\bar q =0, \qquad c=1 \label{PO},\ee corresponding to the
null Killing direction $\hat K=\partial _u .$ It leads to
solutions which depend explicitly on the light-like coordinate
$v=2^{-1/2}(z-t)$ The functions $F$ and $P$ of the Kerr theorem
take in this case the simple form

\be F= -ia Y + (\z - Yv) \ , \ P=1. \label{Fphot} \ee

Solution of the equation $F=0$ yields

\be Y= \z / (ia+ v) \ , \label{Yph} \ee

and function $Y(x)$ determines the principal null congruence $e^3$
by the relation (\ref{tr}). One sees that congruence has a
non--trivial coordinate dependence and has
 a non-zero expansion $\theta$ and twist $\omega$
determined by

\be Z= \theta +i \omega=(v-ia)/[v^2 + a^2] \ , \label{Z1} \ee

which is determined also by generating function $F$ of the Kerr
theorem, (\ref{tr}),

 \be Z^{-1}= \tilde r = - \d_Y F = v+ia . \label{tr1}
\ee

 One sees that
expansion tends to zero only near the front plane $v=0,$ (or
$z=t,$) where the twist $\omega$ is maximal. In the  vicinity of
the axis z, where $Y\Y \to 0,$ and far from the front plane,
congruence tends to simple form $ e^3=du$ corresponding to pp-wave
solutions.

 For the twist-free case $a=0 ,$ congruence represents gradient of
scalar function $e^3= d(u+ \z\Z /v) ,$ and on the front surface,
$v=0 ,$  it is singular, showing that the front $z=t$ is a
shock-plane. However, it should be emphasized that for the
spinning solutions, $a\ne 0 ,$ front is smooth.

If function $Z$ is determined, electromagnetic field (\ref{Fab})
is defined by two functions
 $A$ and $\gamma $ which satisfy to equation (\ref{E1}).
It is easy to check that function \label{Yphot} obeys the
condition $Y,_3 =0. $ It yields simplification of the operator
(\ref{cD}) which  reduces to $\cD=\d _3.$ As a result, the
equations (\ref{E1}), and (\ref{E2}) take the simple form

\be A,_2 = A,_4=0 ,\label{EL1}\ee

\be A,_3 + \bar Z ^{-1} \gamma ,_2 =0, \ \gamma,_4=0 \ .
\label{EL2}\ee

As it follows from (\ref{Fab}), the function $\gamma$ describes a
null electromagnetic radiation propagating along the Kerr
congruence (direction $e^3$). All the considered in \cite{DKS} and
obtained up to now twisting Kerr-Schild solutions were restricted
by the case the $\gamma=0, $ which corresponds to the stationary
and radiation-free electromagnetic fields. We will also presume
here that $\gamma=0$ and will discuss problem of the solutions
with $\gamma \ne0$ in the next section. Since we have
$Y,_2=Y,_3=Y,_4=0 ,$ the equations (\ref{EL1}), and (\ref{EL2})
reduce now to

\be A,_2=A,_3=A,_4=0 \ ,\ee

which means that $A$ may only be an arbitrary holomorphic function
of the complex function $Y(x).$

The gravitational equations (\ref{G5}), and (\ref{G6}) are also
simplified taking the form

\bea M,_2 =0 , \label{G7}  \\
M,_3 = M,_4=0  \label{G8}\eea

with extra condition that $M$ is real. The unique corresponding
solution is $M=const.=m$ Therefore, in spite of the very
nontrivial form of the congruence, but due to extra property
$Y,_3=0 ,$ the equations turn out to be very simple and easily
solvable. As a result, metric is determined by the Kerr-Schild
ansatz (\ref{gks}) with function $h$ given by \be h=[m v - \frac
12 A(Y)\bar A(\bar Y)] / (v^2+a^2) ,\ee where $Y=\frac
{x+iy}{2^{1/2}(v+ia)},$ and

\be e^3 =du - \frac {x^2+y^2}{2(v^2+a^2)} [dv + 2v(xdx +ydy)
+2a(ydx -xdy)]. \label{ex} \ee

Summarizing, we can represent metric in the form

\bea ds^2 =dx^2 + dy^2 + dz^2 -dt^2 + 2\frac{m v - A(Y)\bar A(\bar
Y)} {v^2+a^2} \times \{du \nonumber \\
- \frac {x^2+y^2} {2(v^2+a^2)} [dv + 2v(xdx +ydy) +2a(ydx
-xdy)]\}^2 .\eea

Electromagnetic field is given by $F_\mn= \Re \cF_\mn ,$ where
 $\cF_\mn = \cF_{ab}e^a e^b ,$ is the self-dual field having the
following nonzero tetrad components \be \cF _{12} = \cF _{34} =
A/[v + i a]^2 ; \ \cF _{31} = - A'_Y/[v^2 + a^2]  , \label{A2}\ee
and \be \cF _{23} =\cF _{14}= \cF _{24}=0 \ . \ee

The solutions with with $A=Y^n , \ n = -1, -2,...$ are singular by
$Y \to 0.$ Similar to `gyrons', they represent the spinning
light-like  beams with singular strings positioned along the
z-axis.  Near this string, $Y\to 0,$ the Kerr congruence takes the
constant direction $e^3=du ,$ and
 solution tends to the well known A. Peres pp-wave solutions
\cite{Per,KraSte}.

Note, that the ordinary Kerr geometry has a quadratic in $Y$ Kerr
function $F(Y).$ As a consequence there are two roots of the
solution $F=0$, two different principal null directions
corresponding to type D of metric and the known twofoldedness of
space with a branch line on the Kerr singular ring. This structure
is retained by a boost, however, in the limiting case $v=c$
light-like solution $F$ is linear in $Y .$ Therefore, the degree
of function $F$ changes by jump when the coefficient $p$ becomes
equal to zero in the ultrarelativistic limit.  This fact shed a
light on the difficulties with the obtaining of the
ultrarelativistic limits, which are related with {\it
non-smoothness} of the limiting procedure \cite{BurMag}. In the
limit $v=c,$ the Kerr twofoldedness disappears by jump and
space-time unfolds. The position of the "negative" sheet of the
Kerr solution corresponds to the negative values of the radial
distance. In the considered example $Re \ \tilde r =
2^{-1/2}(z-t), $ which shows that the "positive" and "negative"
Kerr's sheets are placed on the different half-spaces divided by
the front-surface $z=t$. The "negative" sheet becomes the sheet of
advanced fields placing before the front of solution.

It should be noted that one can alternatively consider solutions
with Killing vector $\hat K=\partial _v ,$ which correspond to the
front $u=const.$ moving in opposite directions $z=const. -t .$
Corresponding parameters \be c=q=\bar q =0, \qquad p=1 \label{p1}
,\ee
 lead to quadratic generating function
\be F= -ia Y - Y(u+Y\Z) \  \label{Fantip}, \ee and to function \be
P=Y\Y. \label{Pantip}\ee However, analysis shows that this case
does not differ from the linear one considered
before\footnote{There are also the simplest cases $F=Y+c ,$ which
correspond to pp-wave solutions.}.

The equation $F(Y)=0$ will have two solutions, and one of them,
$Y=0 ,$
 is  nonphysical one, since it leads to $P=0$ and to infinite value for
the tangent vector to Kerr congruence $k_\m=e^3_\m /P.$

Assuming that $Y\ne 0 ,$ we obtain another solution

\be Y= -  (ia+u) / \Z  \label{Yphot1} \ee

which is related to previous one by the space reflection $P :
(x,y,z)\to -(x,y,z)$ and antipodal inversion $J:  Y \to -1/ \Y .$
The `orientifold' transformation $\Omega= P\cdot J$ plays
important role in the particle physics and in superstring theory
(see for example \cite{Joh}). It appears in the complex Kerr
geometry as a necessary element of the complex Kerr string
\cite{BurOri,BurTwi,Bur}.

Finally, one can easily calculate the five complex quantities
determining the degeneracy of the conformal curvature tensor. By
construction, the Kerr-Schild metrics are algebraically special
and $C^{(4)}=C^{(5)}=0.$

We have also
\par
$C^{(1)}= - h,_{11},$
\par
$C^{(2)}= 2Z h,_{1},$
\par
$C^{(3)}= 2Z [h,_{4} + (\bar Z-Z)h].$

Since  $Y,_3=0$, we obtain  $Z,_2= Z,_1=0 ,$ and in the particular
case  $A=const.$ we obtain also $h,_1= h,_{11}=0$ , and
consequently, $C^{(1)}= C^{(2)}= 0,$ and $C^{(3)}= -m Z^3 + e^2
Z^3 \bar Z \ne 0.$

\section{Discussion}
We described the twisting light-like solutions which are direct
descendants of the Kerr-Newman solution and obtained by a smooth
transfer in the parameters of the generating function of the Kerr
theorem. Contrary to corresponding analogs of the Aichelburg-Sexl
solutions, these solutions are twisting, have a non-zero angular
momentum and a smooth (non-singular) front surface.

The parameters $m$ and $a$ of these  solutions correspond to their
relativistic limiting values, i.e. $m=E$ and $a=J/m ,$ obtained by
the condition of the finite value of angular momentum $J=ma.$
During the light-like limit $v\to c ,$ parameter $a_0,$ radius of
the Kerr singular ring, turns out to be shifted to infinity, and
its role is going to the finite relativistic parameter $a = a_0
\sqrt{1- (v/c)^2} $ which acquires principal geometrical meaning,
parametrizing the generating function $F$ of the Kerr theorem and
defining the twist of the null congruence. Sources of the
light-like solutions turn out to be shifted into complex direction
$z \to z+ia ,$ just like the complex source of the Kerr solution
\cite{BurMag,Bur,BurAxi}.

In the light-like case, the Kerr relation $J=ma$ takes the form
$E=J/a ,$ and if we set  $J=\hbar ,$ assuming that this solution
describes gravitational field of a photon, one obtains that $a$ is
de Broglie wavelength of the corresponding  photon relation
$E=\hbar \nu.$
 Together with the obtained by Carter double gyromagnetic ratio of the
Kerr-Newman solution  and with the obtained therein stringy
structures \cite{Bur,BurOri,BurTwi}, as well as  with the reach
spinor-twistor structure connected to the Dirac equation
\cite{BurTwi,BurDir,BurAxi}, it gives one more evidence to the
relationships between the Kerr geometry and structure of spinning
particles.

The considered here case of linear in $Y$ generating function $F$
is important,  since it is simpler of the quadratic one, which
allows one to expect a progress in the analysis of the case
$\gamma \ne0 $ related with the wave electromagnetic fields
\cite{BurAxi}.

Contrary to the Kerr case, the periodic stringlike structure,
which could provide a resonance frequency for electromagnetic
field, is absent in this case. One can expect, that such structure
could appear in some multidimensional generalizations of the
presented solutions. It seems that such a generalization may be
obtained by unification of the Kerr-Schild and Brinkman metrics.

Interest to these solutions is supported for a few reasons related
with the twistorial approach to the problem of gravitational (and
electromagnetic) interactions of spinning particles. The
corresponding Kerr-Schild treatment of multi-particle solutions
\cite{Multi}  is  based on the Kerr theorem with generating
functions $F$ of different degrees in $Y ,$ including the linear
ones. Twistorial structure of a Kerr-Newman particle is described
by a quadratic in $Y$ generating function $F=F_2 (Y),$ (\ref{F2}),
and this case was investigated in details. It was shown
\cite{DKS,KerWil,BurMag} that quadratic in $Y$ functions
correspond to  isolated spinning particles with arbitrary
position, orientation of spin and (finite) boost. Contrary, the
solutions with generating functions of first degree in $Y,$
$F=F_1(Y)$ have not paid considerable attention before, in spite
of their important physical significance - relations to the
light-like spinning particles. The general case of higher degrees
in $Y ,$ considered in \cite{Multi},  showed that multi-particle
Kerr-Schild solutions lead to a multi-sheeted twistor space which
may be split into simple one- and two-sheeted blocks corresponding
to a set of the light-like and massive particles. In particular,
for the generating functions of third degree, $F_3(Y),$ space-time
will be three-fold and may be considered as a product of the
linear in $Y$ function $F_1(Y),$ given by (\ref{Fphot}), and
quadratic in $Y ,$ function $F_2(Y) ,$ given by (\ref{F2}),

\be F_3(Y)= F_1(Y)\cdot F_2(Y). \ee

Then the equation $F_3=0$ will determine two independent
twistorial structures belonging to different Riemannian sheets of
the function $Y(x)$. One of them corresponds to solution
$Y_1(x^\m), \ x^\m \in M^4 ,$ of the equation $F_1(Y)=0 ,$ and
second one - to a twofold solution $Y_2(x^\m)$ of the equation
$F_2(Y)=0 .$ Although these structures are independent, machinery
of the Kerr-Schild formalism shows that the gravitational and
electromagnetic fields of the particles 1 and 2 interact, forming
a singularity at their {\it common} twistor line, fixed by the set
\cite{Multi} \be \{x^\m: \ Y_1(x^\m)= Y_2(x^\m) \}.\ee  In other
words, solution acquires a pole (propagator) \be \sim \frac
1{Y_1(x^\m) -Y_2(x^\m)},\ee which is exhibited in the form of a
singular null-string beam between the particles 1 and 2.

This consideration is close related to the suggested by Nair
\cite{Nai} and renewed by Witten \cite{Wit}
 twistor-string approach to perturbative gauge theory of
 quantum scattering, where the traditional quantum treatment in momentum
 space gets a natural generalization to twistor space with corresponding
 twistorial generalizations of the wave functions and amplitudes of
 scattering.  The gauge bosons are described by curves of first degree
 in twistor space, and the corresponding plane waves are replaced by
 twistor null planes.

The considered here approach based of Kerr theorem seems to be
more informative, since description of the wave function  by a
single twistor null plane, is replaced by section of a twistor
bundle. We expect that the twistorial theory of scattering based
on the Kerr theorem acquires essential advantages and may
represent a new background for quantum theory  of the gauge and
massive spinning particles.

 Besides, since twistorial description of the wave functions contains
 important coordinate information, it represents a very natural way to
incorporate gravity in quantum theory \cite{BurDir}.

Finally, the light-like beam solutions demonstrate an
universality, appearing in many different areas. In particular,
they appear as beams or pp-waves in gravity
\cite{FPV,Per,BurAxi,Fro}, as pp-strings, Schild strings or chiral
constituents of twistor-string in string theory
\cite{Wit,Nai,BurTwi}, as singular twistor lines by interaction of
spinning particles \cite{Multi}, as well as by excitations of the
rotating black-holes \cite{BEHM}. They possess many remarkable
classical and quantum properties which allow one to suppose their
fundamental role in the processes of interaction and, apparently,
in the structure of vacuum.

\section*{Acknowledgement}
We are thankful to R.Kiehn and G.Volovik for stimulating
communications on the early stages of this work. This work has
been supported by the RFBR grant 07-08-00234.

\end{document}